\documentclass[showpacs,preprintnumbers,amsmath,amssymb,prb]{revtex4}

\usepackage{graphicx}
\usepackage{bm}

\begin{document}

\title{Quantum transitions from superfluid to insulating phases in disordered Bose systems}

\author{A.\ V.\ Syromyatnikov}
\email{asyromyatnikov@yandex.ru}
\author{F.\ D.\ Timkovskii}
\email{philippinho@yandex.ru}
\affiliation{Petersburg Nuclear Physics Institute named by B.P.Konstantinov of National Research Center "Kurchatov Institute", Gatchina 188300, Russia}

\date{\today}

\begin{abstract}

By the example of Heisenberg $d$-dimensional disordered non-frustrated antiferromagnets, we discuss quantum transitions at $d\ge2$ from magnetically ordered (superfluid) to various disorder-induced insulating phases (Bose-glass, Mott-glass, etc.) in Bose systems with quenched disorder. We perform a scaling consideration as well as a discussion based on the hydrodynamic description of long-wavelength excitations and on the assumption that the ordered part of the system shows fractal properties near the transition point. We propose that the scaling ansatz for the singular part of the free energy suggested before for the transition to the Bose-glass phase is applicable also for other transitions if the quenched disorder does not produce a local imbalance in sublattices magnetizations. We show using the scaling consideration that $\eta=2-z$ and $\beta=\nu d/2$, where $\eta$, $\beta$, and $\nu$, are critical exponents of the correlation function, the order parameter, and the correlation length, respectively, and $z$ is the dynamical critical index. These relations were missed in previous analytical discussions of Bose-glass and Mott-glass phases. They signify, in particular, that $z=d/2$ for the transition to the Mott-glass phase and that the density of states of localized excitations shows a superuniversal (i.e., independent of $d$) behavior near the transitions. Being derived solely from the scaling analysis, the above relations for $\eta$ and $\beta$ are valid also for the transition to the random-singlet phase.

\end{abstract}


\maketitle

\section{Introduction}

The effect of quenched disorder on properties of quantum systems has attracted much attention in recent years being  probably one of the most fascinating and not completely resolved problems in modern condensed matter physics. \cite{Fisher,sachdev,mirlin,vojta_rev,Zhel13} Quantum magnets and corresponding models of localized spins on a lattice have served as a convenient playground for experimental and theoretical investigations of disordered Bose systems. \cite{vojta_rev,Zhel13} Counterparts of the magnetically ordered and disordered phases in pure isotropic magnetic systems are superfluid (SF) and Mott insulating (MI) states in pure bosonic systems. Influence of quenched disorder on SF-MI quantum phase transitions (QPTs) in non-frustrated Heisenberg antiferromagnets (HAFs) described by the Hamiltonian
\begin{equation}
\label{ham}
 {\cal H} = \sum_{\langle i,j\rangle} J_{ij}\mathbf{S}_i \mathbf{S}_j - h\sum_i S_i^z
\end{equation} 
has been a subject of great interest in recent years, where $\langle i,j\rangle$ denote nearest-neighbor sites of a hypercubic $d$-dimensional lattice. Two transitions have been actively discussed. First, the transition in strong magnetic field $h$ to fully saturated states which belongs to the universality class of Bose-Einstein condensation and which is characterized by dynamical critical exponent $z=2$. Second, the transition in zero magnetic field to magnetically disordered (singlet) states upon varying the exchange coupling strength between couples of nearest spins (dimers). Such a transition is characterized by $z=1$.

Bond-disordered \cite{Yu,yu2} and site-diluted \cite{ros,2d6,2d9,2d10} HAFs \eqref{ham} at $h\ne0$ appears to be a realization of the famous dirty-boson problem considered in detail in Ref.~\cite{Fisher}, where it was shown that non-superfluid Griffiths Bose-glass (BG) phase always intervenes between the ordered (SF) and the paramagnetic (MI) states. Disorder in exchange couplings can produce at zero field another insulating Griffiths phase, Mott-glass (MG) one, between SF and MI states. \cite{2dmg2,2dmg3,2dmg4,2dmg5,2dmg6,2dmg7,2dmg9} MG was observed before in Bose-systems with \cite{Yu,1dmg2,2d7,2dmg1,2dmg8,2dmg10} and without \cite{1dmg00} the particle-hole symmetry. BG and MG states have finite and zero magnetic uniform susceptibility (which corresponds to the compressibility in bosonic systems), respectively. 
\footnote{
The term "Bose glass" was suggested in Ref.~\cite{Fisher} because this disorder-induced state in bosonic systems is a close analog of the Fermi-glass phase proposed by P.W.~Anderson in disordered Fermi systems (see Ref.~\cite{Fisher} for extra discussion). The term "Mott glass" appeared due to the incompressibility of this disorder-induced state (as the Mott insulating state).
}
Another kind of disorder-induced phase arises in HAF on the square lattice upon inhomogeneous bond disorder when probabilities of changing the exchange constant differ for vertical and horizontal bonds. \cite{inbond1,dilut4} Such kind of disorder produces a local imbalance in sublattices magnetizations in antiferromagnetically correlated part of the system that results in the infinite susceptibility of this Griffiths phase. While it is clear that the transition to this disorder-induced state away from the lattice percolation threshold is of a novel universality class, it has not been discussed yet in detail. 

The SF-BG transition was discussed experimentally in suitable magnets in magnetic field \cite{Huv1,wulf,Yu,yama}. We are not aware of experimental consideration of the SF-MG transition despite suitable materials are known ((Tl$_{1-x}$K$_x$)CuCl$_3$ \cite{tlkcucl} and (C$_4$H$_{12}$N$_2$)Cu$_2$(Cl$_{1-x}$Br$_x$)$_6$ \cite{phcc2}) in which this QPT can occur under pressure.

It was noted many times before (see e.g., Refs.~\cite{dilut4,haas,2dmg1,Niederle_2013,percbg}) that transitions from the ordered to disorder-induced phases mentioned above resemble the percolation transition: the ordered part of the system having the form of an infinite network breaks up into domains upon approaching QPT which are surrounded by disordered regions (see Sec.~\ref{app} for extra discussion). Then, main ideas of the theory of collinear magnets on depleted lattices near the percolation threshold \cite{aha,percrev,harkir,shenderfm,shenderaf} were used in Refs.~\cite{we,img} to describe the SF-BG and the SF-MG transitions at $d\ge2$. Those considerations were based on the hydrodynamic description of the long-wavelength spin excitations and on the assumption that the magnetically correlated part of the system shows self-similar (fractal) properties. It was shown in Refs.~\cite{we,img} that results obtained by this approach are fully consistent with the scaling theory suggested in Ref.~\cite{Fisher} for the description of the SF-BG transition. The fractal properties of the ordered part of the system were also explicitly demonstrated in Ref.~\cite{2dmg1} using the strong disorder renormalization group method in 2D systems near the SF-MG transition.

In the present paper, we propose in Sec.~\ref{dilhaf} a unified theory of transitions from the SF to disorder-induced phases at $d\ge2$ (including those mentioned above) which is based on the hydrodynamic description of long-wavelength excitations and on the fractal properties of the ordered part of the system. We find critical behavior of the transition temperature (for $d>2$), the density of states (DOS), and useful relations between critical exponents. 

In Sec.~\ref{qcp}, we discuss briefly the scaling theory based on the simplest ansatz for the singular part of the free energy which was originally proposed in Ref.~\cite{Fisher} for the SF-BG transition and which is suitable for transitions to disorder-induced phases in which disorder does not produce a local imbalance in sublattices magnetizations (e.g., for the SF-BG and the SF-MG transitions). We show using only the scaling theory that
\begin{eqnarray}
\label{eta2}
\eta &=& 2-z,\\
\label{beta2}
\beta &=& \frac{\nu d}{2}
\end{eqnarray}
in quantum systems, where $\eta$, $\beta$, and $\nu$ are critical exponents of the correlation function, the order parameter, and the correlation length, respectively. Surprisingly, Eqs.~\eqref{eta2} and \eqref{beta2} were not derived in Ref.~\cite{Fisher} for $d\ge2$ (only the inequality $2-2d<\eta\le2-z$ was suggested which is valid for all $d\ge1$) and they were also omitted in Ref.~\cite{img} in the consideration of the SF-MG transition. 

We discuss our general findings in Sec.~\ref{disc} and demonstrate that results of Sec.~\ref{dilhaf} are fully consistent with the scaling theory considered in Sec.~\ref{qcp}. We reveal a superuniversal (i.e., independent of $d$) behavior of the DOS in the SF-BG and in the SF-MG transitions. We show that Eq.~\eqref{beta2} gives $z=d/2$ for the SF-MG transition (which should be contrasted with $z=d$ result for the SF-BG transition \cite{Fisher}). We point out that Eq.~\eqref{beta2} is in agreement with numerical results obtained for the SF-BG transition on extra-large 3D systems whereas various smaller magnetic systems studied before both experimentally and numerically show a universal behavior with much smaller $\beta$ values. \cite{critkis,kisel} Then, it was proposed in Ref.~\cite{we} that there is a crossover in the critical region between two critical regimes so that the regime described in Ref.~\cite{Fisher} is realized in the near vicinity of the SF-BG transition (that requires very large systems to reach it). This proposition is supported by the fact that values of the correlation length exponent were found to be the same in small and large systems. Possibly, insufficiently large system sizes in previous numerical considerations of the SF-MG transition are also in the origin of the discrepancy between available numerical findings, Eq.~\eqref{beta2}, and our prediction that $z=d/2$. Then, we hope that our present consideration will stimulate further numerical activity in this field.

It appears somewhat counterintuitive that Eq.~\eqref{beta2} is valid also for bond-diluted quantum systems near the percolation transition (e.g., for HAF bilayer in which randomly chosen couples of nearest spins from different layers are removed): $\beta$ is much larger than the percolation exponent which controls the size of the lattice infinite network signifying that in contrast to previous expectations the long-range magnetic order arises not in the whole volume of the lattice infinite percolating cluster.

Seemingly, the scaling theory proposed in Ref.~\cite{Fisher} is suitable also for the transition from ordered to the random-singlet (RS) phase (which is not a Griffiths phase). \cite{rsp1,rsp2} The latter was discovered recently in disordered 2D $J$-$Q$ model in which apart from the Heisenberg exchange the multi-spin non-frustrated interaction $Q$ is introduced. \cite{rsp1,rsp2} Equality \eqref{eta2} derived below using only the scaling arguments was really observed in Refs.~\cite{rsp1,rsp2} numerically.

Sec.~\ref{conc} contains a summary and the conclusion.

\section{Qualitative discussion of transitions to considered disorder-induced Griffiths phases}
\label{app}

\begin{figure}
\centering
\includegraphics[scale=0.7]{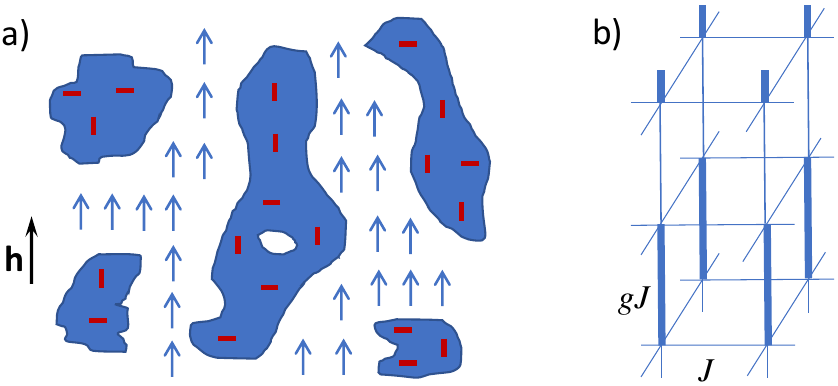}
\caption{(a) Sketch of system \eqref{ham} near the SF-BG transition. Bonds with increasing exchange constants are shown in red. Regions are shaded with non-saturated local magnetization. (b) Three-dimensional version of model \eqref{ham} in which SF-MG transition arises at $h=0$ upon variation of $g$ values on randomly chosen bold bonds. $J>0$ and $gJ>0$ are exchange constants.
\label{sfbgfig}}
\end{figure}

In order to illustrate the qualitative similarity of considered transitions from the ordered state to glassy phases, we discuss first in some detail model \eqref{ham} assuming that the magnetic field is large. In pure system, the transition takes place at $h=h_{c0}$ from the magnetically ordered (SF) state with broken continuous symmetry and Goldstone excitations to the magnetically disordered (MI) fully saturated phase with unbroken continuous symmetry and with gapped spectrum. The order parameter is the mean transverse spin component in this transition. Let us introduce the disorder to this system by increasing exchange constants $J_{ij}$ on a small number of randomly chosen bonds. To saturate spins around defect bonds, a field greater than $h_{c0}$ is required. Then, local values are greater of the saturation field in regions with more dense distribution of defects. As a result, one comes to the following qualitative picture. All spins have finite mean transverse components at $h<h_{c0}$. At $h>h_{c0}$, regions saturate with smaller local critical fields and the magnetically ordered part of the system looks like an infinite network. Ordered regions leave the infinite network before $h$ approaches their local critical field values if they are surrounded by areas with lower local critical fields. Such regions do not contribute to the net order parameter of the system which is given by the order parameter of the infinite network. The latter falls to clusters of finite volume at a critical field $h_c>h_{c0}$ which is the SF-BG transition point (see Fig.~\ref{sfbgfig}(a)). Thus, this transition is qualitatively similar to the percolation transition as it was noted in many papers (see, e.g., Refs.~\cite{dilut4,haas,2dmg1,Niederle_2013,percbg}).

Similar physical picture arises in model \eqref{ham} at zero field upon bond randomness mentioned in the Introduction. Let us consider for definiteness the system shown in Fig.~\ref{sfbgfig}(b) with exchange coupling constants $J>0$ and $gJ>0$. The transition takes place in this system from the N\'eel ordered state to the disordered phase with singlet (dimerized) ground state at $g=g_{c0}>1$. Let us increase and fix $g$ values at a small amount of randomly chosen bold bonds shown in Fig.~\ref{sfbgfig}(b). One expects that spin pairs near a defect bond become dimerized at $g$ values smaller than $g_{c0}$. Thus, local critical values of $g$ would be smaller in regions with a more dense distribution of defects. Similar to the transition to the BG phase, the N\'eel ordered part of the system aquires the form of an infinite network, the number of sites in it decreases upon $g$ increasing, and it disappears (falls to clusters of finite volume) at $g=g_c<g_{c0}$. The order parameter of this transition is the mean staggered magnetization. Because the volume of a finite cluster is not bounded from above, all considered glassy phases have gapless spectra. 

It is the fractal (self-similar) lattice geometry at the percolation threshold that makes universal the percolation transition. Then, it is assumed in Sec.~\ref{dilhaf} that universality of considered transitions from the SF to Griffiths phases has the same origin: fractal properties of the magnetically correlated part of the system which differ only quantitatively from those of the percolation fractal. (It should be mentioned that the fractal properties of the ordered part of the system were explicitly demonstrated in Ref.~\cite{2dmg1} using the strong disorder renormalization group method in 2D systems near the SF-MG transition.) Thus, it is natural to consider models \eqref{ham} using methods applied before in Refs.~\cite{percrev,harkir,shenderaf,shenderfm} to collinear diluted HAFs near the percolation threshold. This consideration is presented in Sec.~\ref{dilhaf}.

In further general discussion, we denote $g$ the control parameter driving the system to the phase transition and its value at the transition will be designated as $g_c$.

\section{Ordered states near transitions to disorder-induced Griffiths phases}
\label{dilhaf}

Hydrodynamic excitations, which we assume to exist, can be described phenomenologically using the following expression for the system energy in the continuum limit: \cite{harkir,shenderaf,shenderfm}
\begin{equation}
\label{e}
E = \int d{\bf r} 
\sum_{\alpha=1}^{d-1}
\left(
\frac{\Upsilon}{4M^2} 
\left(
\overrightarrow{\nabla} m_1^\alpha({\bf r}) - \overrightarrow{\nabla} m_2^\alpha({\bf r})
\right)^2
+
\frac{1}{2\chi_\perp} \left( m_1^\alpha({\bf r}) + m_2^\alpha({\bf r})\right)^2
\right),
\end{equation}
where ${\bf m}_1({\bf r})$ and ${\bf m}_2({\bf r})$ are transverse components of sublattices magnetizations, $M$ is the staggered magnetization per unit volume, $\Upsilon$ is the helicity modulus, and $\chi_\perp$ is a uniform susceptibility in an infinitesimal field transverse to the staggered magnetization. Eq.~\eqref{e} is valid both at zero and at finite magnetic field on length scales larger than the correlation length $\xi$. 

Spectrum of the hydrodynamic excitations found from the Landau-Lifshitz equations for Fourier components of ${\bf m}_1({\bf r})$ and ${\bf m}_2({\bf r})$ has the form \cite{harkir,shenderaf,shenderfm}
\begin{equation}
\label{specaf}
	\epsilon_{\bf k} = 
	\sqrt{\frac{2\Upsilon}{\chi_\perp}} k.
\end{equation}
These propagating excitations arise near the quantum critical point (QCP) on the length scale greater than $\xi\propto|g_c-g|^{-\nu}$. On smaller length scale, excitations are expected to be localized as in diluted HAFs \cite{percrev,Fisher,we,img} and they are called "fractons".
\footnote{
As soon as percolation transition, SF-BG, and SF-MG transitions are qualitatively similar (see above), we adopt some terminology from the theory of the percolation transitions, in which localized high-energy excitations are commonly named "fractons" \cite{percrev,aha}. The term "fracton" here should not be confused with the emergent topological quasiparticle with restricted mobility in fracton phases belonging to recently discovered new quantum state of matter \cite{fractons}.
}

We assume also the sharp phase transition and the Harris criterion fulfillment in the disordered systems with the conventional power-law scaling. Quantities introduced above scale near $g_c$ as
\begin{eqnarray}
\label{m}
M &\propto& (g_c-g)^{\beta},\\
\label{ups}
\Upsilon &\propto& (g_c-g)^{\sigma},\\
\label{chi}
\chi_\perp &\propto& (g_c-g)^{-\gamma}.	
\end{eqnarray}
Notice that the transverse susceptibility $\chi_\perp$ behaves differently near the considered transitions. It remains constant (i.e., $\gamma=0$ in Eq.~\eqref{chi}) near the transition to BG state \cite{Fisher} because the system is already in magnetic field (so that an application of an infinitesimal magnetic field ${\bf h}_\perp$ transverse to sublattices magnetizations produces a magnetization in that direction proportional to $h_\perp$). 
$\chi_\perp$ scales near the SF-MG transition as the volume of the infinite network in which the long-range antiferromagnetic order exists (i.e., $\gamma=-\beta<0$) because the rest part of the system is in a singlet state with a gapped spectrum. \cite{img} $\chi_\perp$ diverges (i.e., $\gamma>0$) in systems with the disorder-induced local imbalance in sublattices magnetizations (randomly diluted HAFs and HAFs with the inhomogeneous random bond dilution). The latter is due to uncompensated net spins arising on large length scales which produce a large magnetization of the whole system in an infinitesimal external field (see Refs.~\cite{harkir,percrev} for more details). The distinct behavior of $\chi_\perp$ produces the difference between formulas derived below for considered transitions.

The DOS of the infinite network carrying the long-range order can be found as in diluted HAFs (see Refs.~\cite{harkir,shenderaf,percrev}). The result has the form
\begin{eqnarray}
\label{dosinf}
{\cal D}_{inf}(\omega) 
&\propto &
\left\{
\begin{aligned}
&(g_c-g)^{\beta}\omega^{\upsilon_1}, \quad && 
\omega \gg \omega_0 \sim (g_c-g)^{\nu+(\sigma+\gamma)/2} = (g_c-g)^{\nu z},	\\
&\frac{\omega^{d-1}}{(g_c-g)^{d(\sigma+\gamma)/2}}, \quad && \omega\ll\omega_0,	
\end{aligned}
\right.\\
\label{v1}
\upsilon_1 &=& \frac{d-z}{z} - \frac{\beta}{\nu z},\\
\label{z}
z &=& 1 + \frac{\sigma+\gamma}{2\nu},
\end{eqnarray}
where $z$ is the dynamical critical exponent. The DOS of propagating excitations \eqref{specaf} gives the second line in Eq.~\eqref{dosinf}. Eq.~\eqref{specaf} at $k\sim1/\xi$ gives the characteristic energy $\omega_0$. Taking into account that properties of excitations (energy and the DOS normalized to one spin) on the spatial length scale smaller than $\xi$ (and at energies $\omega\gg\omega_0$) are independent on $g_c-g$ (i.e., on the proximity to the QCP), \cite{shenderaf,halph} one comes to the first line in Eq.~\eqref{dosinf}. The first line in Eq.~\eqref{dosinf} matches the second line at $\omega\sim\omega_0$ if Eq.~\eqref{v1} holds. Using the geometrical relation \cite{foot} 
\begin{equation}
\label{geom}
	\beta = (d-D)\nu,
\end{equation}
where $D<d$ is the fractal dimension, one can also simplify Eq.~\eqref{v1}.

The N\'eel temperature $T_N(g)$ can be estimated at $d>2$ as it was done in Ref.~\cite{shenderaf} for the diluted HAF near the percolation transition. We introduce for that an auxiliary quantity (which has no transparent physical meaning and introduced solely for the sake of $T_N(g)$ estimation)
\begin{equation}
\label{phidef}
\Phi(k) = \frac{E(k)}{\left|{\bf m}_1\left({\bf k}\right)\right|^2},
\end{equation}
where $E(k)$ is energy \eqref{e} of the spin-density wave characterized by momentum $\bf k$. At $k\ll1/\xi$, one can find $\Phi(k)$ from Eq.~\eqref{e} by expressing ${\bf m}_2({\bf k})$ via ${\bf m}_1({\bf k})$ using Landau-Lifshitz equations and substituting the result to Eq.~\eqref{e}. As a result, one obtains $\Phi(k) = \epsilon_{\bf k}^2\chi_\perp/2M^2$, where $\epsilon_{\bf k}$ is given by Eq.~\eqref{specaf}. Because the energy $\omega$ of the spin-density wave characterized by momentum $\bf k$ is given by $\epsilon_{\bf k}$ at $k\ll1/\xi$, one finds from this result $\Phi(\omega)=\omega^2\chi_\perp/2M^2$ at $\omega\ll\omega_0$. As $\Phi(\omega)$ is proportional to the system excitation energy, it does not depend on $g_c-g$ at $\omega\gg\omega_0$ (see Refs.~\cite{shenderaf,halph} and above). Thus, one can try it in the form $\omega^{\upsilon_\phi}$ at $\omega\gg\omega_0$, where $\upsilon_\phi$ has to be found by matching expressions for $\Phi(\omega)$ in these two regimes at $\omega\sim\omega_0$. As a result, we have
\begin{eqnarray}
\label{phi}
\Phi(\omega) 
&\propto &
\left\{
\begin{aligned}
& \omega^{\upsilon_\phi}, \quad && \omega\gg\omega_0,	\\
& \omega^2 (g_c-g)^{-\gamma-2\beta}, \quad && \omega\ll\omega_0,	
\end{aligned}
\right.\\
\upsilon_\phi &=& 2-\frac{\gamma+2\beta}{z\nu}.
\end{eqnarray}
The reduction of sublattices magnetization $\delta M$ due to thermal fluctuations at $T\ll T_N(g)$ and $g<g_c$ is estimated using Eq.~\eqref{phidef} as \cite{shenderaf}
\begin{equation}
\label{dmm}
\frac{\delta M}{M}
\sim \frac{\langle m_1^2\rangle}{2M^2}
=
\sum_{\bf k} \frac{\langle |{\bf m}_1({\bf k})|^2\rangle}{2M^2}
=
\frac{1}{2M^2} 
\sum_{\bf k} \left\langle \frac{E(k)}{\Phi(k)} \right\rangle,
\end{equation}
where $\langle \dots \rangle$ denotes the thermal average. The summation over $\bf k$ in the right-hand side of Eq.~\eqref{dmm} is actually a summation over excited states of the system which can be standardly replaced by the integration over energy involving the DOS, the result being \cite{shenderaf}
\begin{equation}
\label{dm}
\frac{\delta M}{M} = \frac{1}{2M^2} 
\int\frac{\omega{\cal D}_{inf}(\omega)}{(e^{\omega/T}-1)\Phi(\omega)}d\omega,
\end{equation}
where the thermal averaging gives the factor $1/(e^{\omega/T}-1)$. It can be demonstrated using Eqs.~\eqref{dosinf}, \eqref{phi}, and \eqref{dm} that $\delta M\ll M$ at $T\ll\omega_0$. At larger temperatures, $\omega_0\lesssim T\ll T_N(g)$, the main contribution to the integral in Eq.~\eqref{dm} comes from $\omega\sim T$ and $\omega\sim\omega_0$ when $\beta +\gamma+\nu(d-2z) > 0$ and $\beta +\gamma+\nu(d-2z)<0$, respectively. As a result, Eq.~\eqref{dm} gives $\delta M/M \sim (g_c-g)^{-\beta} T^{(\beta+\gamma+\nu(d-z))/z\nu} $ and $\delta M/M \sim T (g_c-g)^{\nu(d-2z)+\gamma}$ in these two cases, correspondingly. These expressions should be valid by the order of magnitude at $\delta M\sim M$ and $T\sim T_N(g)$ that allows to obtain the following estimation for the N\'eel temperature:
\begin{eqnarray}
\label{tn}
T_N(g) &\propto& (g_c-g)^{\phi},\\
\label{phi1}
\phi &=& 
\left\{ 
	\begin{aligned}
	&\frac{\beta z\nu}{\beta+\gamma+\nu(d-z)}, && \beta +\gamma+\nu(d-2z) \ge 0,\\
	&\nu(2z-d)-\gamma, && \beta +\gamma+\nu(d-2z) \le 0.
	\end{aligned}
\right.
\end{eqnarray}

Let us consider the DOS of finite clusters ${\cal D}_{fin}(\omega)$. Our assumptions about the qualitative similarity of the transitions under discussion with the percolation transition implies that finite clusters of all characteristic linear sizes smaller than $\xi$ appear near the QCP and that the probability to find larger clusters is exponentially small (see Refs.~\cite{aha,percrev}). As a consequence, ${\cal D}_{fin}(\omega)=0$ at $\omega\ll\omega_0$. As the number of sites in finite clusters does not vanish at $g=g_c$, we try ${\cal D}_{fin}(\omega)$ in the form $\omega^{\upsilon_2}$ at $\omega\gg\omega_0$, where $\upsilon_2$ can be found as follows. We point out first that the DOS per spin in the infinite network is of the order of $\omega_0^{\upsilon_1}$ at $\omega\sim\omega_0$ (see Eq.~\eqref{dosinf}). The DOS per spin has the same form in the largest clusters having the characteristic linear size of the order of $\xi$. \cite{aha,percrev} As a result, one estimates
$
V_\xi\omega_0^{\upsilon_1} \sim {\cal D}_{fin}(\omega\sim\omega_0) \propto \omega_0^{\upsilon_2}
$, 
where $V_\xi$ is the total volume of the largest clusters. $V_\xi$ is of the order of the volume of the infinite network $(g_c-g)^\beta$ because the latter falls predominantly into clusters with the linear size of the order of $\xi$ \cite{aha,percrev}. One obtains as a result
\begin{eqnarray}
\label{dosfin}
{\cal D}_{fin}(\omega) 
&\propto &
\left\{
\begin{aligned}
&\omega^{\upsilon_2}, \quad &&\omega\gg\omega_0,	\\
&0, \quad &&\omega\ll\omega_0,
\end{aligned}
\right.\\
\label{v2}
\upsilon_2 &=& \frac{d-z}{z}.
\end{eqnarray}

In particular, one derives for the specific heat
\begin{eqnarray}
\label{c}
{\cal C}  &\propto &
\frac{d}{dT}\int\frac{\omega{\cal D}(\omega)}{e^{\omega/T}-1}d\omega
\end{eqnarray}
at $g=g_c$ using Eqs.~\eqref{dosfin} and  \eqref{v2}
\begin{equation}
\label{c2}
	{\cal C} \sim T^{d/z}.
\end{equation}

The long-distance behavior of the equal-time correlation function has the form at $g=g_c$ \cite{Fisher}
\begin{equation}
\label{chir0}
	\chi(r)\sim r^{-(d+z-2+\eta)}.
\end{equation}
According to our assumptions, $\chi(r)$ should be equal to the fractal correlation function ${\cal G}(r)\sim r^{-2(d-D)}$ which gives the probability that two sites located at a distance $r$ from each other belong to the same cluster. \cite{foot} Using also Eq.~\eqref{geom}, one comes as a result to the standard hyperscaling relation
\begin{equation}
\label{eta}
\eta = \frac{2\beta}{\nu}+2-d-z.
\end{equation}

Formulas obtained in this section does not rely on any assumptions concerning the scaling behavior of the free energy and space-time correlators of the order parameter. They should be valid in the entire critical region and should describe also crossovers (if any) from one critical behavior to another (see also discussion in Ref.~\cite{we}). However much more detailed predictions can be made if the scaling ansatz for the free energy is known. We are not aware of such ansatz for systems in which the disorder produces local imbalance in the sublattices magnetization (e.g., percolation transitions in cite-diluted magnets and systems with the inhomogeneous bond depletion). However, the ansatz is known for systems with the disorder-preserved local sublattices balance in which transitions arise to the BG or to the MG phases. We consider this ansatz in the next section and show in Sec.~\ref{disc} that all predictions made in the present section are fully consistent with that scaling consideration.

\section{Scaling theory}
\label{qcp}

It is proposed in Ref.~\cite{Fisher} that the SF-BG transition is described by the following simplest scaling ansatz for the singular part of the free energy:
\begin{equation}
\label{ansatz}
	f_s(g,T)\sim |g_c-g|^{\nu(d+z)} F \left( \frac{T}{(g_c-g)^{\nu z}}, \frac{\tilde h}{|g_c-g|^{\nu (d+z)-\beta}} \right),
\end{equation}
where $\tilde h$ is the field conjugated to the order parameter. It is also shown in Ref.~\cite{dilut1} that Eq.~\eqref{ansatz} is applicable to the percolation transition in bond-diluted HAFs (e.g., in HAF bilayer in which randomly chosen couples of nearest spins from different layers are removed). It was proposed in Ref.~\cite{img} that the derivation in Ref.~\cite{dilut1} of Eq.~\eqref{ansatz} is applicable to SF-MG transitions under the assumption about the fractal nature of the ordered part of the system near the QCP. Then, we extend this statement now to all systems in which disorder does not produce local magnetic moments because the derivation of Eq.~\eqref{ansatz} proposed in Ref.~\cite{dilut1} remains valid under this condition (see also discussion in Refs.~\cite{dilut2,dilut3,hoyos}).

A detailed quite general consideration of ansatz \eqref{ansatz} was proposed in Ref.~\cite{Fisher}. Then, we refer the reader to that paper for extra details and present below only relations from Ref.~\cite{Fisher} which are necessary for explanation of our new findings in this field.

It follows from Eq.~\eqref{ansatz} that the compressibility $\kappa$ scales as $\kappa \sim (g_c-g)^{\nu(d-z)}$ (generalized Josephson relation). \cite{Fisher} Because the uniform susceptibility \eqref{chi} is the counterpart of the compressibility in HAFs, we obtain 
\begin{equation}
\label{z2}
z = d+\frac\gamma\nu.
\end{equation}

The N\'eel temperature \eqref{tn} is derived from Eq.~\eqref{ansatz} by noting that the function $F(y,0)$ must be singular at some point $y=y_c$ in order for the system to show the transition at some temperature $T_N(g)$. \cite{Fisher} Then, one has $T_N(g)=y_c(g_c-g)^{z\nu}$ and
\begin{equation}
\label{phi2}
\phi = z\nu.
\end{equation}

It follows also from Eq.~\eqref{ansatz} that the specific heat 
\begin{equation}
\label{heat}
	{\cal C} =-T \frac{\partial^2 f_s}{\partial T^2} \propto T^{d/z}
\end{equation}
at $g=g_c$. \cite{Fisher} Bearing in mind relation \eqref{c} of the specific heat with DOS, it follows from Eq.~\eqref{heat} that DOS scales at the QCP as 
\begin{equation}
\label{dos2}
	{\cal D}(\omega)\propto \omega^{(d-z)/z}.
\end{equation}

Scaling \eqref{ansatz} implies also that the long-distance and the long-time behavior of the order-parameter dynamical susceptibility has the form near the QCP
\begin{equation}
\label{chirt}
	\chi(r,\tau) 
	= \overline{\left \langle T_\tau a({\bf r},\tau)a^\dagger({\bf 0},0) \right\rangle}
	- \overline{\langle a({\bf r},\tau)\rangle \langle a^\dagger({\bf 0},0)\rangle} 
	\sim r^{-(d+z-2+\eta)} w \left( \frac r\xi, \frac{\tau}{\xi^z} \right),
\end{equation}
where lines denote the disorder average, $a({\bf r},\tau)$ is the Bose operator, and $w(x,y)$ is an appropriate function (in particular, $w(0,0)={\rm const}$ so that Eq.~\eqref{chirt} is consistent with Eq.~\eqref{chir0}). \cite{Fisher} Eq.~\eqref{chirt} gives at $g=g_c$ for the long-time correlator \cite{Fisher} 
\begin{equation}
\label{chi0t}
	\chi(0,\tau) \sim \tau^{-(d+z-2+\eta)/z}.
\end{equation}
On the other hand, $\chi(0,\tau)$ is equal at the QCP to the averaged Green's function 
$$
\overline{G({\bf r=0},\tau) }
	= \overline{\left \langle T_\tau a({\bf r=0},\tau)a^\dagger({\bf 0},0) \right\rangle}
$$
which is related to the DOS as
\begin{equation}
\label{gf}
	\overline{G({\bf r=0},\tau) }
	= -{\rm sign}(\tau) \int_0^\infty d\omega e^{-\omega|\tau|} {\cal D}({\rm sign}(\tau)\omega),
\end{equation}
where ${\cal D}(\omega)$ should scale as
\begin{equation}
	\label{dos3}
	{\cal D}(\omega) \propto \omega^{(d-2+\eta)/z}
\end{equation}
in order Eq.~\eqref{gf} to be consistent with Eq.~\eqref{chi0t}. Comparing Eqs.~\eqref{dos2} and \eqref{dos3} on comes to Eq.~\eqref{eta2} and the hyperscaling relation \eqref{eta} (which can be obtained \cite{Fisher} from Eqs.~\eqref{ansatz} and \eqref{chirt}) gives in turn Eq.~\eqref{beta2}. Surprisingly, Eqs.~\eqref{eta2} and \eqref{beta2} were not derived in Ref.~\cite{Fisher} for $d\ge2$ (only the inequality $2-2d<\eta\le2-z$ was suggested which is valid for all $d\ge1$) although Eqs.~\eqref{phi2}, \eqref{heat}, \eqref{chirt}--\eqref{dos3} were presented there. The only new ingredient which we add as compared to the consideration in Ref.~\cite{Fisher} is relation \eqref{c} of the specific heat with DOS which readily leads to Eq.~\eqref{dos2} at the QCP.  

\section{Discussion}
\label{disc}

\subsection{Consistency of the scaling consideration in Sec.~\ref{qcp} with results from Sec.~\ref{dilhaf} at $\gamma\le0$}

The consistency should be stressed between our findings in Sec.~\ref{dilhaf} and the scaling theory proposed in Ref.~\cite{Fisher} and based on ansatz for the singular part of the free energy \eqref{ansatz}. In particular, one derives using Eq.~\eqref{ansatz} for the helicity modulus $\Upsilon\propto (g_c-g)^{\nu(d+z-2)}$ that gives together with Eq.~\eqref{ups} $\sigma=\nu(d+z-2)$. 
\footnote{
Difference $\Delta f_s\propto \Upsilon (\nabla\varphi)^2$ of the free energy arises as a result of imposition of gradient $\nabla\varphi$ of the order parameter phase along a spatial direction. As the scaling dimension of $\nabla\varphi$ is the inverse length, $\nabla\varphi\sim1/\xi$ and $\Upsilon\propto (g_c-g)^{\nu(d+z-2)}$. \cite{Fisher}
} 
Substituting this equality to Eq.~\eqref{z} one obtains Eq.~\eqref{z2}. We come to Eq.~\eqref{phi2} from Eqs.~\eqref{phi1} and \eqref{z2} in both cases of $\beta + \gamma + \nu(d-2z) \le 0$ and $\beta + \gamma + \nu(d-2z) \ge 0$. Hyperscaling relation \eqref{eta} is derived in Ref.~\cite{Fisher} from Eq.~\eqref{ansatz} and scaling arguments not using the fractal correlation function. Eq.~\eqref{heat} is in agreement with Eq.~\eqref{c2}. Eq.~\eqref{dos2} agrees with Eqs.~\eqref{dosfin} and  \eqref{v2}.

\subsection{Site-diluted and inhomogeneously bond-diluted antiferromagnets. $\gamma>0$.}

The local imbalance in sublattices magnetizations produced by disorder in site-diluted HAFs near the percolation threshold and in the inhomogeneously bond-diluted HAFs near and away from the percolation threshold results in the divergence of the uniform susceptibility upon approaching the QCP from the ordered side ($\gamma>0$). \cite{harkir,percrev,dilut4,inbond1} Our discussion in Sec.~\ref{dilhaf} is a generalization of results obtained in Refs.~\cite{shenderaf,harkir} (see also the review Ref.~\cite{percrev}) for diluted HAFs near the percolation threshold. In particular, one finds $\phi=\sigma-\nu<z\nu$ (cf.\ Eq.~\eqref{phi2}) at $d=3$ using values of percolation critical exponents and Eq.~\eqref{phi1} that signifies inapplicability of ansatz \eqref{ansatz}. The origin of this inapplicability is in the disorder-induced local imbalance in sublattices magnetizations. \cite{dilut2,dilut3,hoyos} It was found \cite{percrev} that localized excitations (fractons) in the infinite cluster show a superuniversal behavior at $d\ge2$ near the percolation threshold:
\begin{equation}
\upsilon_1=0
\end{equation}
(see Eqs.~\eqref{dosinf}).

To the best of our knowledge, the critical behavior has not been discussed yet in inhomogeneously bond-diluted HAFs away from the percolation threshold. It is clear, however, that the transition in this system should be of a novel universality class which is described by formulas from Sec.~\ref{dilhaf}.

\subsection{Bose-glass phase. $\gamma=0$.}

The SF-BG transition was discussed in detail in the seminal paper Ref.~\cite{Fisher}, where it was shown, in particular, that 
\begin{equation}
\label{zbg}
z=d.	
\end{equation}
We supplemented that consideration in Ref.~\cite{we} with Eqs.~\eqref{eta2} and \eqref{beta2} and with the observation that the DOS per spin of fractons both in the infinite network and in isolated clusters shows the superuniversal behavior: 
\begin{equation}
\label{dosbg}
	\upsilon_1=-1/2, \quad \upsilon_2=0
\end{equation}
(see Eqs.~\eqref{v1} and \eqref{v2}). Available numerical estimations of critical exponents in 2D systems are close to Eqs.~\eqref{eta2}, \eqref{beta2}, and \eqref{zbg}. \cite{2d10,2d9,2d6,2d7,2d1,2d2} 

We compare in Table~\ref{table2} our predictions for critical exponents at $d=3$ with available experimental and numerical findings. It was shown in Ref.~\cite{kisel} that very large 3D systems (with up to $2\times20^6$ sites) are required to reach the crossover to the critical behavior predicted in Ref.~\cite{Fisher}. In particular, $\beta=1.5(2)$ was found numerically in Ref.~\cite{kisel} in agreement with Eq.~\eqref{beta2} (see Table~\ref{table2}). The value of $\beta=1.08(20)$ obtained in Ref.~\cite{dtnx} for system with up to $60^3$ sites is also in agreement with Eq.~\eqref{beta2}. In contrast, the universal behavior with smaller values $\beta\sim 0.6-0.9$ (and $\phi<\nu d$) were observed on various much smaller magnetic systems \cite{Yu,yu2,haas,critkis,kisel} (see also discussion in Ref.~\cite{we}). While numerical data for $\eta$ and $\beta$ obtained in Refs.~\cite{hitch,haas,yu2} in 3D systems are in agreement with Eqs.~\eqref{eta2} and \eqref{beta2}, $\phi<\nu d$ was found in Refs.~\cite{haas,yu2} and it was pointed out in Ref.~\cite{hitch} that larger systems are needed for a reliable determination of $\eta$.

\begin{table}
\caption{Critical exponents for the SF-BG transition at $d=3$ obtained experimentally and numerically on systems containing $L^3$ sites with $L$ up to $L_{max}$. Values of critical exponents are also indicated given by  Eqs.~\eqref{eta2}, \eqref{beta2}, \eqref{phi2}, and \eqref{zbg} with $\nu=0.88(5)$ (which was observed in Ref.~\cite{kisel} in the largest systems).
\label{table2}
}
\begin{ruledtabular}
\begin{tabular}{|c|ccccc|}
   & Eqs.~\eqref{eta2}, \eqref{beta2}, \eqref{phi2}, \eqref{zbg} & & numerics, & numerics, & experiment \\
 & with $\nu=0.88(5)$ & & $L_{max}\ge60$ & $L_{max}\le20$ & \\
\hline
$\nu$ 					& --- & & $0.88(5)$ [\cite{kisel}]; 0.75(11) [\cite{dtnx}] & 0.75(10) [\cite{Yu,yu2,critkis}]; 0.7(1) [\cite{haas,hitch}]; 0.9(1) [\cite{percbg}] & --- \\
$\phi$ 		& $2.6(2)$ & & $2.7(2)$ [\cite{kisel}] & 1.1(1) [\cite{Yu,yu2,critkis,percbg}]; 1.16(5) [\cite{haas}] & $1.1-1.2$ [\cite{Yu,Huv1,yama}]\\
$z$ 					& 3 & & 3 [\cite{kisel,dtnx}] & 3 [\cite{Yu,yu2,haas,critkis,hitch}] & --- \\
$\beta$ & $1.32(8)$ & & 1.5(2) [\cite{kisel}]; 1.08(20) [\cite{dtnx}] & 0.6(1) [\cite{kisel}]; 0.95(10) [\cite{Yu,yu2}]; 0.9(1) [\cite{haas}]; 1.2(1) [\cite{percbg}] & $0.4-0.5$ [\cite{Huv1,wulf}]\\
$\eta$			& $-1$ & & --- & $\approx-1$ [\cite{haas,Yu,yu2,hitch}] & ---
\end{tabular}
\end{ruledtabular}
\end{table}

\subsection{Mott-glass phase and bond-diluted systems near the percolation threshold. $\gamma=-\beta<0$.}

The SF-MG transition was discussed in the spirit of the present paper in Ref.~\cite{img}, where, however, Eqs.~\eqref{eta2} and \eqref{beta2} were overlooked somehow. Taking into account Eq.~\eqref{geom}, one concludes from Eq.~\eqref{z2} that $z$ is equal to the fractal dimension $D$. \cite{img} Eq.~\eqref{beta2} makes the expression for $z$ even more definite:
\begin{equation}
\label{zmott}
	z=d/2.
\end{equation}
Similar to the transitions in pure systems, the dynamical exponent in the SF-MG transition \eqref{zmott} is two times smaller than that in the SF-BG transition \eqref{zbg}. One observes also from Eqs.~\eqref{v1} and \eqref{v2} using Eqs.~\eqref{zmott} and \eqref{beta2} the superuniversal behavior of fractons 
\begin{equation}
\label{dosmg}
	\upsilon_1 = 0, \quad \upsilon_2=1.
\end{equation}

We point out that the fulfillment of the generalized Josephson relations (for the superfluid density and the compressibility) was demonstrated numerically in Ref.~\cite{lerch} for the SF-MG transition in 2D systems that implies the fulfillment of scaling ansatz \eqref{ansatz} (see Ref.~\cite{Fisher}) and, consequently, Eqs.~\eqref{eta2}, \eqref{beta2}, and \eqref{zmott}. Then, the equality $z=d-\beta/\nu$ which follows from Eq.~\eqref{z2} for the SF-MG transition is in good agreement with results of recent numerical considerations of 2D \cite{2dmg3,2d7,2dmg8,dilut2,dilut3,lerch} and 3D \cite{sfmg3d} systems. However all the available numerical results for $z$, $\eta$, and $\beta$ deviate from Eqs.~\eqref{eta2}, \eqref{beta2}, and \eqref{zmott} as it is illustrated by Table~\ref{tablemg}. The situation here may be similar to that with the SF-BG transition described above, where very large 3D systems were required to achieve the crossover to the critical regime predicted in Ref.~\cite{Fisher}.
\footnote{
Notice that in the case of the crossover, consideration performed in Sec.~\ref{dilhaf} could be valid in the whole critical region.
}
On the other hand, the analysis in Ref.~\cite{2dmg1} of the superfluid part of the 2D system performed within the strong disorder renormalization group method gave $z$ values slightly larger than 1 (in good agreement with Eq.~\eqref{zmott}) and values of $\eta$ and $\beta$ were fully consistent with Eqs.~\eqref{eta2} and \eqref{beta2} (see Table~\ref{tablemg}).

\begin{table}
\caption{Critical exponents for the SF-MG transition at $d=2$ and 3 obtained numerically, within the strong disorder renormalization group (SDRG) method (only for $d=2$) \cite{2dmg1}, and using Eqs.~\eqref{eta2}, \eqref{beta2}, and \eqref{zmott}.
\label{tablemg}
}
\begin{ruledtabular}
\begin{tabular}{|c|cccc|ccc|}
	 &  & $d=2$  &  &   & & $d=3$ & \\
	& Eqs.~\eqref{eta2}, \eqref{beta2}, \eqref{zmott} & & SDRG & numerics 
		& Eqs.~\eqref{eta2}, \eqref{beta2}, \eqref{zmott} & & numerics\\
\hline
$z$         & 1 & & 1.31(7) [\cite{2dmg1}] & 1.52(3) [\cite{2dmg8}]; 1.5(2) [\cite{2d7}]
   & 3/2 & & 1.67(6) [\cite{sfmg3d}]\\
$\beta/\nu$ & 1 & & 1.1(2) [\cite{2dmg1}] & 0.48(2) [\cite{2dmg8}]; 0.60(15) [\cite{2d7}] 
& 3/2 & & 1.09(3) [\cite{sfmg3d}]\\
$\eta$			& 1 & & 0.9(2) [\cite{2dmg1}] & $-0.52(4)$ [\cite{2dmg8}]; $-0.3(1)$ [\cite{2d7}]
& 1/2 & & $-0.50(3)$ [\cite{sfmg3d}]
\end{tabular}
\end{ruledtabular}
\end{table}

\begin{figure}
\centering
\includegraphics[scale=0.7]{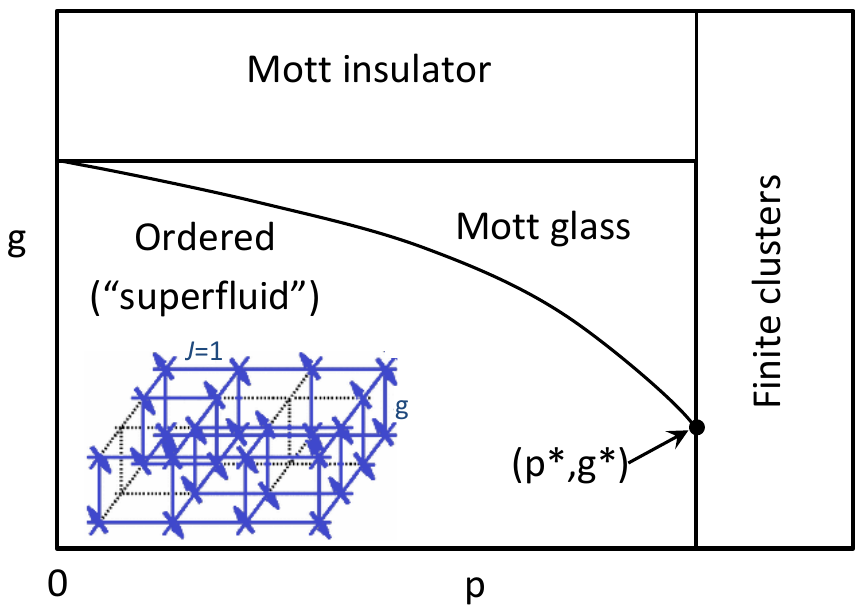}
\caption{Sketch of the phase diagram of antiferromagnetic bilayer in which randomly chosen couples of nearest spins from different layers are removed (see the inset). \cite{dilut5,dilut1,2dmg5,2dmg9,2dmg3,2dmg6,hoyos,Puschmann} Here $p$ is the concentration of discarded couples and $g$ controls the exchange coupling of the remainder couples. The multicritical point $(p^*,g^*)$ is depicted, where $p=p^*$ is the percolation threshold of the lattice.
\label{pd}}
\end{figure}

It should be noted that the scaling consideration in Sec.~\ref{qcp} should be valid also in dimer-diluted systems near the percolation threshold. The example of such system is HAF bilayer in which randomly chosen couples of nearest spins from different layers are removed (see the inset in Fig.~\ref{pd}). It is the system for which ansatz \eqref{ansatz} was derived analytically in Ref.~\cite{dilut1}. An intuitively clear idea was proposed in Ref.~\cite{dilut1} that $\beta$ should be equal to the percolation exponent of the lattice infinite cluster volume. However Eq.~\eqref{beta2} resulting from the analysis of quantum correlators predicts much larger $\beta$ value. Although previous numerical works in diluted quantum HAFs on a single spin-1/2 layer \cite{sand,percrev} and on the bilayer \cite{2dmg5} near the percolation threshold obtained $\beta$ values consistent with its percolation value, the order parameter exponent can differ in these two systems: there are a lot of extensive one-dimensional elements in the lattice percolation infinite cluster \cite{percrev} in which antiferromagnetic correlations are expected to decay algebraically (as in spin-1/2 chains) and exponentially (as in two-leg spin ladders) in the spin-1/2 single-layer and in the bilayer quantum systems, respectively. According to these simple considerations, one expects that the long-range magnetic order arises not in the whole volume of the infinite cluster of the lattice in the quantum bilayer that is consistent with our result \eqref{beta2}.

The phase diagram of HAF bilayer obtained numerically before is presented in Fig.~\ref{pd}. \cite{dilut5,dilut1,2dmg5,2dmg9,2dmg3,2dmg6,hoyos,Puschmann} Critical behavior was distinguished numerically along the line $p=p^*$, in the vicinity of the multicritical point $(p^*,g^*)$, and away from this point at $p<p^*$ (critical exponents were found to be different in these three regimes). Interestingly, Eqs.~\eqref{eta2} and \eqref{zmott} predict the same $z$ and $\eta$ for all three regimes.

Thus, we hope that our findings will stimulate further numerical discussion of the SF-MG transition.

\subsection{Random-singlet phase. $\gamma=0$.}

The SF-RS transition was observed recently numerically in 2D $J$-$Q$ model and it was proposed that scaling ansatz \eqref{ansatz} is relevant to this transition and even to the RS (critical) state. \cite{rsp1,rsp2} It was found that the uniform susceptibility remains finite in the RS phase signifying $\gamma=0$ as in the BG phase. It was obtained also that $z=d$ and $\eta=2-z$ at the QCP in agreement with Eqs.~\eqref{zbg} and \eqref{eta2}.

\section{Summary and conclusion}
\label{conc}

To conclude, by the example of disordered $d$-dimensional non-frustrated Heisenberg antiferromagnets we discuss quantum phase transitions at $d\ge2$ from the ordered (superfluid) to disorder-induced insulating phases which qualitatively differ from each other by the scaling of the uniform susceptibility (compressibility) \eqref{chi}: Bose-glass ($\gamma=0$), Mott-glass ($\gamma=-\beta<0$), and the Griffiths phase in the inhomogeneously bond-diluted HAFs ($\gamma>0$). We perform consideration which is valid for all these phases and which is based on the hydrodynamic description of the long-wavelength spin excitations and on the assumption of the self-similar (fractal) nature of the ordered part of the system. 

We propose that the scaling theory which is based on ansatz \eqref{ansatz} for the free energy and which was proposed before for the SF-BG transition \cite{Fisher} and for homogeneously bond-diluted systems near the percolation threshold \cite{dilut1} is suitable also for systems in which disorder does not produce a local imbalance in sublattices magnetizations. Using only the scaling arguments, we find Eqs.~\eqref{eta2} and \eqref{beta2} for $d\ge2$ which were omitted in previous considerations of the SF-BG \cite{Fisher} and the SF-MG \cite{img} transitions. These relations lead to the prediction $z=d/2$ for the SF-MG transition (see Eq.~\eqref{zmott}) which should be contrasted with $z=d$ (see Eq.~\eqref{zbg})  obtained in Ref.~\cite{Fisher} for the SF-BG transition. Interestingly, in bond-diluted quantum systems near the percolation threshold, Eqs.~\eqref{eta2}, \eqref{beta2}, and \eqref{zmott} remain valid and the order parameter critical exponent $\beta$ given by Eq.~\eqref{beta2} is much larger than its percolation value signifying that the long-range order arises in these quantum systems not in the whole volume of the lattice infinite cluster. Because this our finding contradicts available numerical results and because the applicability of scaling ansatz \eqref{ansatz} to homogeneously bond-diluted systems near the percolation threshold was proven analytically \cite{dilut1}, further numerical activity is required in this field. Eqs.~\eqref{eta2} and \eqref{beta2} lead also to the superuniversal (i.e., independent of $d$) behavior of DOS of localized excitations (see Eqs.~\eqref{dosinf}, \eqref{dosfin}, \eqref{dosbg}, and \eqref{dosmg}).

Notice that only numerical findings obtained in very large systems near SF-BG transition are consistent with Eq.~\eqref{beta2} (as well as with Eq.~\eqref{phi2}) whereas results obtained on smaller systems give a universal behavior with much smaller $\beta$ values and $\phi<\nu z$. \cite{critkis,kisel} Then, it is proposed in Ref.~\cite{we} that there is a crossover in the critical region between two critical regimes so that the regime described in Ref.~\cite{Fisher} is realized in the near vicinity of the SF-BG transition (that requires very large systems to reach it). We suppose that similar situation arises in 2D SF-MG transition and insufficiently large system sizes are in the origin of the deviation of previous numerical results for 2D SF-MG transition from our predictions \eqref{beta2} and \eqref{zmott}. Thus, we hope that the present paper will stimulate further experimental and numerical activity in this field.

It was proposed in Refs.~\cite{rsp1,rsp2} that scaling ansatz \eqref{ansatz} is applicable also to the transition from the ordered to the random-singlet phase (which is not a Griffiths phase) in 2D systems. Numerical findings in Refs.~\cite{rsp1,rsp2} are in agreement with our prediction \eqref{eta2} made solely from the scaling analysis.

\begin{acknowledgments}

This work is supported by the Russian Science Foundation (Grant No.\ 22-22-00028) and by the Foundation for the Advancement of Theoretical Physics and Mathematics "BASIS".

\end{acknowledgments}

\bibliography{Unifbib}

\begin{thebibliography}{61}
\bibitem[{foo()}]{foot}
\bibinfo{howpublished}{One can obtain this equality as it is done in
  percolation theory. \cite{aha,percrev}}.
\end{thebibliography}

\end{document}